# Ultracompact Blue Dwarf Galaxies: *Hubble Space Telescope* Imaging and Stellar Population Analysis

Michael R. Corbin[1,2], William D. Vacca[3], Roberto Cid Fernandes[4], John E. Hibbard[5],

Rachel S. Somerville[6], & Rogier A. Windhorst[1]

_______________________________________________________________________

[1]Department of Physics & Astronomy, Arizona State University, P.O. Box 871504, Tempe, AZ 85287;

 Rogier.Windhorst@asu.edu

[2]Current address: U.S. Naval Observatory, Flagstaff Station,  10391 W. Naval Observatory Rd., Flagstaff,

 AZ, 86001-8521; mcorbin@nofs.navy.mil

[3]SOFIA-USRA, MS 211-3, NASA/Ames Research Center, Moffett Field, CA

 94035-1000; wvacca@mail.arc.nasa.gov

[3]Departmento de Física – CFM – Universidade Federal de Santa Catarina, Florianópolis, SC, Brazil;

cid@astro.ufsc.br

[4]National Radio Astronomy Observatory, 520 Edgemont Road, Charlottesville, VA 22903;

jhibbard@nrao.edu

[5] Max Planck Institut für Astronomie, Königstuhl 17, D-69117, Heidelberg, Germany;

somerville@mpia.de

ABSTRACT

We present deep *Hubble Space Telescope* Advanced Camera for Surveys / High Resolution Channel *U*-band, narrow *V*-band, and *I*-band images of nine "ultracompact" blue dwarf galaxies (UCBDs) selected from the Sloan Digital Sky Survey (SDSS). We define UCBDs as local ($z < 0.01$) star-forming galaxies having angular diameters less than $6''$ and physical sizes less than 1 kpc. They are also among the most metal-poor galaxies known, including objects having $12 + \log(O/H) < 7.65$, and are found to reside within voids. Both the *HST* images and the objects' SDSS optical spectra reveal that they are composites of young (~1 Myr) populations that dominate their light, and older (~10 Gyr) populations that dominate their stellar masses, which we estimate to be $\sim 10^7 - 10^8 \, M_\odot$. An intermediate age ($\sim 10^7 \, \mathrm{yr} - 10^9 \, \mathrm{yr}$) population is also indicated in most cases. The objects are not as dynamically disturbed as the prototype UCBD, POX 186, but the structure of several of them suggests that their current starbursts have been triggered by the collisions/mergers of smaller clumps of stars. In one case, HS 0822+3542, the ACS/HRC images resolve the object into two small (~100 pc) star-forming components which appear to have recently collided, supporting this interpretation. In six of the objects much of the star formation is concentrated in Young Massive Star Clusters, contributing to their compactness in ground-based images. The evidence that the galaxies consist mainly of ~10 Gyr old stars establishes that they are not protogalaxies, forming their first generation of stars. Their low metallicities are more likely to be the result of the escape of supernova ejecta, as opposed to youth. The evidence that old stellar populations dominate the masses of the objects is consistent with recent galaxy formation simulations which predict that cosmic re-ionization at $z \cong 6$ significantly limited the subsequent star formation of dwarf galaxies in voids due to the photo-evaporation of baryons from their cold dark matter halos.

*Subject Headings*: galaxies: formation --- galaxies:dwarf --- galaxies: starburst --- galaxies: stellar

content

1.  INTRODUCTION

Blue compact dwarf galaxies (BCDs, also referred to as H II galaxies) have long been suspected of being young objects because of their vigorous star formation and low metallicities (e.g. Searle & Sargent 1972; see also the review by Kunth & Östlin 2000). However, nearly all BCDs have been found to consist of starburst regions within envelopes of significantly evolved (ages of several Gyr) stars, and are thus not newly formed galaxies (e.g. Raimann et al. 2000; Kong et al. 2003; Westera et al. 2004). Recent studies of the most metal-poor BCDs including I Zw 18 have produced conflicting results regarding the presence of evolved (ages greater than ~500 Myr) stars in these objects, leaving open the question of whether *some* BCDs indeed represent objects forming their first generation of stars (see Izotov & Thuan 2004 and Momany et al. 2005 for I Zw 18; Östlin & Mouhcine 2005; Corbin et al. 2005; Aloisi et al. 2005; Izotov, Thuan, & Guseva 2005; Pustilnik, Kniazev, & Pramskij 2005, for other objects). Both possibilities have important implications for understanding galaxy formation, as newly formed galaxies in the local universe would allow the formation process to be studied in greater detail than in objects at very high redshifts, while the presence of large numbers of evolved stars in all local dwarf galaxies would establish that they formed nearly contemporaneously with their more massive counterparts. In either case, BCDs deserve careful study because they represent the extreme low ends of the galaxy mass, luminosity, and metallicity functions, and thus place basic constraints on models of galaxy formation.

Imaging of BCDs has revealed a variety of morphologies (e.g. Loose & Thuan 1985; Kunth, Maurogordato, & Vigroux 1988; Papaderos et al. 1996; Telles, Melnick, & Terlevich 1997; Cairos et al. 2001; Gil de Paz, Madore, & Pevunova 2003). The majority have sizes, shapes, and underlying red stellar populations similar to quiescent dwarf elliptical galaxies, suggesting an evolutionary connection between these classes (e.g., Telles et al. 1997; but see van Zee, Salzer, & Skillman 2001 for possible complications to a simple evolutionary model). Other BCDs have more irregular structure, including "cometary" objects with elongated shapes and highly asymmetric distributions of their star formation (see, e.g. Loose & Thuan 1985; Noeske et al. 2000; Izotov & Thuan 2002; Aloisi et al. 2005). The study of Kunth et al. (1988) revealed one BCD, POX 186, to be "ultracompact," with angular and physical sizes only ~3″ and ~ 300 pc, respectively. Their images also showed little evidence of an underlying population of evolved stars. Subsequent ground-based imaging of POX 186 at subarcsecond resolution by Doublier et al. (2000) and with the *Hubble Space Telescope* by Corbin & Vacca (2002) confirmed the extremely small physical size of this object, and also revealed a strong asymmetry suggestive of a recent collision between two even smaller (~100 pc and ~$10^7$ M$_\odot$) clumps of stars that might be regarded as subgalactic in nature. This result suggests that POX 186 is a dwarf galaxy in the process of formation, or at least *assembly*, given that the colors of the outer regions of the object indicate the presence of evolved stars (Doublier et al. 2000; Corbin & Vacca 2002).

The POX 186 results strongly motivate *Hubble Space Telescope* (*HST*) imaging of more ultracompact blue dwarf galaxies (hereafter UCBDs). We thus searched Data Release 2 of the Sloan Digital Sky Survey (SDSS) to select a homogeneous sample of UCBDs, taking as defining criteria an SDSS $g$-band angular diameter $d < 6″$, as measured from the Petrosian radii tabulated in the SDSS (which are measured with a Petrosian ratio of 0.2), a physical diameter $D < 1$ kpc, and a spectroscopic redshift $z < 0.01$, to image them at high spatial resolution. These selection criteria represent arbitrary limits to the angular and physical size distribution of the BCD population, designed to identify objects that are only marginally resolved from the ground, and which are thus good targets for *HST*. The possible physical differences between UCBDs and BCDs is one of the issues to be addressed in this paper.

Our search revealed that UCBDs are rare, with only nine objects in Data Release 2 of the SDSS meeting the above criteria. While no color limit was imposed, all of the objects meeting the selection criteria were found to be actively star-forming, appearing uniformly blue in the SDSS color composite images (with no clear evidence of an evolved population). They also have spectra containing strong emission lines. Our definition of a UCBD is thus less restrictive than the most recent definitions of BCDs based on color (e.g. Gil de Paz et al. 2003), but the basic criteria are effectively the same. Several of these nine objects are also classified as "extremely metal poor" by Kniazev et al. (2003), having $12 + \log(O/H) < 7.65$ as measured from the emission lines in their SDSS spectra. This further motivates *HST* imaging, in order to resolve their structure and test for the presence of evolved stars. Our selection criteria allow the objects be resolved to a physical scale ~10 pc with the High Resolution Channel of the Advanced Camera for Surveys (hereafter ACS/HRC) of *HST*.

In this paper we present the ACS/HRC images and photometry of these nine UCBDs, along with an analysis of their stellar populations from their SDSS optical spectra. Our principal finding from both the images and spectra is that all of the objects have an underlying evolved stellar population with an age of at least ~10 Gyr, along with recent (~1 – 10 Myr old) starbursts. Stars of intermediate age (~$10^7 – 10^9$ yr) are also indicated in most of the objects. Several of the objects (as well as POX 186) also contain young massive star clusters (YMCs), and possibly "super" star clusters (SSCs), similar to those found in other BCDs (e.g. Thuan, Izotov, & Lipovetsky 1997; Johnson et al. 2000). SSCs and YMCs have been found in both major mergers and non-interacting galaxies, and the former may represent newly formed globular clusters (see, e.g., O'Connell, Gallagher, & Hunter 1994; Whitmore 2000; Larsen 2004). Our ACS/HRC images of one of the sample objects, HS 0822+3542, clearly resolve it into two small and apparently interacting components that are arguably subgalactic in nature, and those data are discussed in greater detail in a separate paper (Corbin et al 2005; hereafter C05). The morphologies of some of the remaining objects are suggestive of recent mergers/collisions between smaller clumps of stars that may have triggered the objects' current starbursts.

In the following section we describe the sample properties and the *HST* observations. In § 3 we present the images, spectra, and their analysis. We conclude with a discussion of the results in § 4. A cosmology of $H_0 = 71$ km s$^{-1}$ Mpc$^{-1}$, $\Omega_\Lambda = 0.73$, $\Omega_M = 0.27$ is assumed (Spergel et al. 2003).

## 1. SAMPLE PROPERTIES AND HST OBSERVATIONS

### 2.1 Sample Properties

The basic properties of the nine sample objects are summarized in Table 1. The *g*-band magnitudes and redshifts are from the SDSS database[1]. Our distance estimates represent averages of four distance values based on the objects' redshifts and the Virgocentric infall corrections of Schechter (1980), Huchra (1988), Binggeli, Popescu, & Tammann (1993), and Mould (1995). We also give the distance to the nearest neighboring galaxy, as determined from the NASA Extragalactic Database (NED), which includes all objects discovered in the SDSS. These distances, along with inspection of the SDSS images, show that all of the objects are isolated. The three objects that have a neighbor within 1 Mpc still reside in very sparse environments, and the neighboring galaxies are dwarfs. For example, the object SDSS J0825+3532 (= HS 0822+3542) resides in a void, and its nearest neighbor is a low surface brightness dwarf galaxy; the

[1]see http://www.sdss.org

nearest large galaxy lies approximately 3 Mpc away (Pustilnik et al. 2003). BCDs are also found in low-density environments (Campos-Aguilar, Moles, & Masegosa 1993; Telles & Terlevich 1995; Popescu,

Hopp, & Rosa 1999; Telles & Maddox 2000).

<div align="center">2.2 *HST Observations and Photometry*</div>

The ACS/HRC has a pixel scale of 0.027″ pix$^{-1}$, corresponding to a physical scale ~10-20 pc per nominal resolution element at the distances of the objects. We obtained deep images of all nine objects in the sample in the F330W, F550M ("narrow *V*"), and F814W filters of the ACS/HRC. These filters cover nearly the complete wavelength range of the ACS/HRC, and were chosen to avoid the strong emission lines in the objects' spectra including [O III] $\lambda\lambda$4959, 5007 and H$\alpha$. The measured fluxes should thus mainly represent the stellar continuum emission of the objects, with a contribution from nebular continuum emission depending on the strength and age of the starburst (see Leitherer & Heckman 1995; Kruger, Fritze-von Alvensleben, & Loose 1995). We estimate the contribution of this nebular component from the objects' SDSS spectra in § 3.3. The dates of the observations and the integration times in each filter are presented in Table 2. The images were corrected for Galactic foreground extinction using the maps of Schlegel et al. (1998) before creating the color images and measuring magnitudes.

Table 3 presents the photometric properties of each object. The object magnitudes were measured from the drizzled and sky-subtracted images produced by the *HST* reduction pipeline using the STSDAS.ELLIPSE task (the magnitudes for components A and B of SDSS J0825+3532 were measured using polygonal apertures; see C05). Conversion from count rates to fluxes was made using the conversion factors in the image header files. The first column gives the length in arcseconds of the major axis of the elliptical aperture in which the object's light was measured. The second column gives the limiting surface brightness of the F814W image, as measured from the sky level beyond this aperture. The remaining columns give the object magnitude and colors, and the absolute magnitudes in the F550M filter calculated from the distances given in Table 1. The $3\sigma$ limiting AB magnitudes of point sources in the images measured from their background levels are approximately 25.9 (F330W), 25.7 (F550M), and 26.6 (F814W). These limiting magnitudes allow the detection of individual OB and red supergiant stars at the distances to the galaxies, but not individual main-sequence and giant stars of later spectral type.

## 1. RESULTS

### 3.1 *Color Images*

The composite ACS/HRC F330W, F550M, and F814W images of all nine galaxies are shown in Figure 1, where the images have been loaded into the blue, green, and red color channels, respectively. The images are displayed on a logarithmic intensity scale because of the large range in intensity between the starburst and quiescent regions of the objects. The same color table was used in all images. The figures also show the angular size and corresponding physical scale at the adopted distances to the objects. The images are scaled such that they show the structure of the objects down to approximately the limiting surface brightnesses given in Table 1, so in no cases do they represent only the "tip of the iceberg." In addition, a gray-scale image of SDSS J0825+3532 is presented in C05, and confirms that the object consists of two physically separate components, as opposed to high surface regions in a single object. We also tried to photometrically separate the point sources and unresolved stars in the galaxies using the DAOPHOT and SExtractor packages, but these attempts failed due to their complexity.

The images clearly reveal most of the objects (notably SDSS J1119+5130, SDSS J1127+6410, and SDSS J1133+6349) to be composites of populations of different ages, with starburst regions embedded in envelopes of red stars. The decomposition of the stellar populations of the objects is best determined from their spectra, which is the subject of § 3.3. We however find F550M – F814W colors in the areas away from the starburst regions to be in the approximate range 0.7 – 1.8, consistent with populations dominated by red giant stars.

Individual stars are evident in most of the objects, and measurement of the absolute magnitudes and colors of selected stars using the adopted distances to the galaxies confirms them to be OB stars and red supergiants. Concentrated dust is not seen in any cases, nor is strong nebular continuum emission, e.g., in shells or filaments surrounding the starburst regions or the stellar envelopes of the galaxies themselves, with the possible exception of SDSS J0248-0817 (see § 3.2). SDSS J0825+3532 is the only object resolved into two smaller components by the ACS/HRC images. Several of the objects contain YMCs, and we present photometry of them in § 3.5. None of the objects show obvious tidal features of the type seen in POX 186 (Corbin & Vacca 2002), but they are not particularly symmetric either. We discuss each object individually in the following section.

## 3.2 *Comments on Individual Objects*

*SDSS J0133+1342* - This is one of the most distant and intrinsically smallest objects in the sample. It is also one of the bluest, with its light dominated by several possible YMCs (see § 3.4) and also a relatively blue component of unresolved stars. Several red supergiant stars are also visible, establishing an upper limit to the age of the starburst(s) of $\sim 10^8$yr if their progenitors were OB stars and if an instantaneous burst is assumed. SDSS J0133+1342 also qualifies as "extremely metal poor" according to Kniazev et al. (2003), with a value of $12 + \log(O/H) = 7.60 \pm 0.03$. Its morphology is irregular, with a suggestion of a tidal feature in the southeast direction. Its spectral energy distribution (§ 3.3) indicates that an underlying population of evolved stars is present, although it is not clearly seen in the color composite image, possibly as the result of low surface brightness.

*SDSS J0236-0058* – The light from this object is dominated by a single YMC near its center, with individual OB stars and red supergiants also detected. An underlying relatively low surface brightness (approximately 21.5 mag arcsec$^{-2}$ in F814W) red stellar envelope is detected, which appears to be more concentrated towards the north.

*SDSS J0248-0817* – This object qualifies as a "cometary" type of BCD, with a large starburst region at its northern edge and a relatively low surface brightness component (approximately 20.5 mag arcsec$^{-2}$ in F814W) of redder stars extending to the south. Only the starburst region was detected and cataloged in the SDSS object database, leading to our identification of it as a UCBD candidate. While it is not strictly a UCBD under our criterion of a physical size less than 1 kpc, its small size and morphology in the SDSS images was sufficiently striking to warrant inclusion in the sample. The starburst region has a complex structure, consisting of two central loose clusters of stars with surrounding associations and individual OB stars. Two shell-like structures are seen at the northern edge of the starburst region, with recently formed stars on their edges. This may be be an example of supernovae-induced star formation (see, e.g. Oey 2004; Oey et al. 2005).

*SDSS J0825+3532* – This remarkable object has been studied in detail by C05, Pustilnik et al (2003) and Kniazev et al. (2000). The ACS/HRC images reveal it to consist of two very small clumps of stars (designated A and B) which appear to have recently collided, triggering their star formation (C05). A faint plume of redder stars including individual supergiants extends to the northwest of both components, supporting an interaction picture. In this paper we use a different distance to the object of 15.4 Mpc versus 12.7 Mpc (C05) based on the distance estimate method discussed in § 2.1, which results in a slightly different size scale (Figure 1). In C05 we suggest that SDSS J825+3532 represents a dwarf galaxy in the process of assembling from clumps of stars that are subgalactic in terms of size and mass.

*SDSS J1119+5130* - This relatively bright and blue object was discovered by Arp (1965), who described it as a "very small, condensed galaxy," and it is often referred to as "Arp's Galaxy" in the literature. It has also occasionally been referred to as a galaxy pair, because of its marginal separation into two components in ground-based images. The ACS/HRC images reveal these two components to be starburst regions in a single underlying red stellar envelope. Numerous red supergiants are also evident, particularly near the edge of the galaxy. We find one candidate YMC in the western starburst region. Kniazev et al. (2003) classify this object as extremely metal poor, with $12 + \log(O/H) = 7.51 \pm 0.04$.

*SDSS J1127+6410* - This object displays an interesting combination of recent star formation and an underlying component of red stars morphologically similar to a dwarf elliptical. It differs however from the most common type of BCD (see § 1) in that the star formation is more evenly distributed across the galaxy, as opposed to being concentrated near the center, and is not concentrated in clusters.

*SDSS J1133+6349* - This object is very similar to SDSS J1119+5130, consisting of two basically distinct starburst regions and a roughly elliptical underlying redder population. It also contains two YMCs in the eastern starburst region and several red supergiants.

*SDSS J1135+0153* - The recent star formation in this relatively red object is occurring mainly in an approximately linear structure near its northwestern edge, where individual OB stars and red supergiants are seen. The main body of the galaxy is of relatively low surface brightness (approximately 21.5 mag arcsec$^{-2}$ in F814W) compared to other objects in the sample

*SDSS J1152-0040* - The emission of this object is dominated by the light of a YMC at its center, with some star formation also evident at its southwestern tip. Like SDSS J0236-0058, it contains a number of red supergiants and an underlying and low surface brightness (approximately 21.5 mag arcsec$^{-2}$ in F814W) component of redder stars.

### 3.3 *Stellar Population Analysis*

The *HST* images of the sample objects and their colors reveal them to be composites of old and young stellar populations. In order to examine this more precisely, we retrieved their optical spectra from the SDSS database and analyzed them with the STARLIGHT code of Cid Fernandes et al. (2004, 2005; Mateus et al. 2005). STARLIGHT fits an observed spectrum with a combination of multiple simple stellar populations (SSPs; also known as instantaneous burst) synthetic spectra using a $\chi^2$ minimization procedure. The specific SSP spectra chosen for our analysis are from Bruzual & Charlot (2003), which are based on the STELIB library of Le Borgne et al. (2003), Padova 1994 evolutionary tracks, and a Chabrier (2003) Initial Mass Function between 0.1 and 100 M$_\odot$. We use the same set of 25 ages used by Mateus et al. (2005) for their analysis of galaxies in SDSS Data Release 2. We adopt a metallicity of $Z = 0.0004$ (= 1/50 $Z_\odot$) for all models, which is the metallicity in the Bruzual & Charlot (2003) library closest to the metallicities of the sample galaxies based on their O/H abundances (Kniazev et al. 2003). We use an SMC-reddening law to treat internal extinction because of the similarity of the sample objects to the SMC. The spectra were also corrected for Galactic extinction using the maps of Schlegel, Finkbeiner & Davis (1998), shifted to the rest-frame, and re-sampled to a resolution of 1 Å. Bad pixels and emission lines were excluded from the final fits. The spectra of SDSS J0133+1342 and SDSS J0825+3532 were too noisy for STARLIGHT to run successfully. The spectral energy distributions of the two components of SDSS J0825+3532 as constrained by their fluxes in the F330W, F550M, and F814W filters are presented by C05, where it was found that both components appear to be composites of at least two distinct stellar populations with ages ~1 Myr and ~10 Gyr.

The fits obtained from the STARLIGHT code are presented in Figure 2. The SDSS spectral fibers are 3″ in diameter, and thus include either all or the majority of the light in each of the sample objects (see Fig. 1). In the case of the cometary galaxy SDSS J0248-0817, the spectrum represents the starburst region at its northern tip. In addition to the Bruzual & Charlot (2003) simple stellar population synthesis models, we include a nebular continuum spectrum to assess its possible contribution to the observed spectra (see § 2.2). This nebular continuum is calculated for an electron temperature of $10^4$ K and assumes $n$(He II)/ $n$(H I) = 0.1 and $n$(He III)/ $n$(H I) = 0.0, with emission coefficients from Aller (1984). We verified that the calculated nebular continuum strengths are in good agreement with the strengths expected on the basis of the objects' H$\beta$ fluxes. We find in all objects however that the contribution of this continuum component does not exceed 10%, and is below 1% in several objects. This is consistent with the finding of C05 for SDSS 0825+3542. The observed continuum in all the objects is therefore predominantly stellar. The spectral decomposition is illustrated by grouping the SSP spectra into four age bins plus the nebular component, whose strengths are listed in each plot. In each object we find the best fit to be obtained by a combination of young ($\leq 10^7$ yr) , intermediate-age (~$10^7$ yr – $10^9$ yr), and old ($\geq 5 \times 10^9$ yr) populations. The age of the older populations are found to be closest to approximately 10 Gyr. The combined uncertainties in the models and data actually limit the age resolution to ~0.5 – 1.0 dex, but this does not affect the conclusion (in qualitative agreement with the images) that stellar populations spanning several decades in age are present in *all* of the objects.

In Figure 3 we present the best fit of the $Z$ = 0.0004 Bruzual & Charlot (2003) models to the total

F330W, F550M, and F814W fluxes of SDSS 0133+1342, since its SDSS spectrum is too noisy for the

STARLIGHT code to reliably fit. The fit was based on six models with ages between 1 My and 10 Gyr,

and was obtained by a simple $\chi^2$ minimization procedure, not STARLIGHT. Although the fit is not as

strongly constrained as those from the SDSS spectra, we obtain consistent results, in that the presence of at

least a very young (~1 Myr) and an old (~10 Gyr) population is indicated in the object, as in SDSS

0825+3532 (see C05). A single-age population cannot fit the ACS/HRC filter fluxes, and the overall spectral energy distribution can only be fit by the inclusion of a population with an age of 1 Gyr or older.

The extended structure of SDSS J0248-0817 (Fig. 1) allows us to examine the spectral energy distribution of the region near its geometric center using the ACS/HRC images. We specifically measure the flux in an aperture 1.62″ square, centered at pixels x = 615, y = 547, in all three images. This central region allows us to more directly measure the underlying red population of the galaxy than the starburst region at its tip. Fitting the fluxes with the same set of SSP models applied to SDSS 0133+1342, we find that the best fit to the fluxes is obtained from the combination of a 1 Gyr old and a 10 Gyr old population.

This is shown in Figure 4. This result is qualitatively consistent with the fit to the starburst region spectrum

(Figure 2), although there is a significant difference in the relative contributions of such populations in these regions. The spectral energy distribution of this central region nonetheless cannot be fitted with a population younger than ~1 Gyr.

### 3.4 *Stellar Masses*

Stellar masses of the sample objects were estimated from the final spectral fits, using luminosities calculated from the adopted distances to the galaxies (Table 1), and mass-to-light ratios calculated from each objects' population mixture (*M/L* values are ~1 for the *z*-band). A correction for light outside of the 3″ SDSS spectroscopic fiber was made by comparing the *z*-band magnitudes of the objects measured inside the fiber to their total *z*-band magnitudes. In addition, the regions of the galaxies covered by the fibers will likely be the bluest, and will have different *M/L* ratios than the regions outside of the fibers, which affects the estimates of their total masses. To correct for this, we estimated the *z*-band *M/L* ratios of the regions outside of the spectroscopic apertures using their F550M – F814W colors, applying calibrations from a large sample of galaxies selected from the SDSS. Details of this procedure are given in Appendix A. For SDSS J0248-0817, the SDSS *z*-band magnitude includes only the northern starburst region, so our mass estimate is a lower limit.

The mass values are given in Table 4, and represent the mass currently contained in both stars and stellar remnants, which are included in the Bruzual & Charlot (2003) models. Table 4 also gives the fraction of the stellar mass contained in the young, intermediate-age, and evolved stellar populations within the spectroscopic apertures, and the *V*-band stellar extinction values obtained from the fitting procedure. The total mass values we obtain are very small, but are consistent with those found from the photometric data on POX 186 (Corbin & Vacca 2002) and HS 0822+3542 (C05), which assumed a total mass-to-luminosity ratio of ~2 – 4 (Thuan 1987). We note that component B of HS 0822+3542 has an estimated mass of ~2 – 4 $\times 10^6$ M$_\odot$, smaller than the mass of the sample objects and approaching that of a globular cluster. It may however be representative of the masses of the clumps of stars whose collisions/merging may have triggered the objects' star formation, as was suggested for POX 186 by Corbin & Vacca (2002).

Figure 5 shows the distribution of relative masses the different stellar populations, as determined from the spectral fitting procedure. It also shows the relative contributions of these populations to the light of the objects at 4020 Å. This plot and the values in Table 3 show that the ~10 Gyr-old stars in the objects comprise ~75-98% of their mass, while the young and intermediate-age stars in the bluer objects together contribute a similar fraction of the light at that wavelength. The uncertainties in the mass fractions in the three age bins used (Table 4) are approximately 6% (Cid Fernandes et al. 2005). Inferring the star formation histories of the objects from these mass distributions is, however, not simple, given the necessary binning of the stellar masses (see, e.g., Mathis, Charlot, & Brinchmann 2006). These distributions also only represent the light inside of the spectroscopic fibers; the fractions of light and mass represented by older stars will increase if the total galaxies are considered, given that the regions away from the galaxy are redder (see Appendix A).

*3.5  Young Massive Star Clusters*

Several of the UCBDs contain objects that appear to be YMCs, taking as selection criteria a marginally resolved structure and an absolute magnitude $M_{F550M} < -9$ mag. We present a list of these candidates and associated photometry in Table 5. The magnitudes were measured with the IRAF "phot," task, using circular apertures whose radii varied from 3 to 6 pixels. Aperture corrections were applied using the plots and tables of Sirianni et al. (2005). A background level was measured and subtracted using circular annuli of width 2 pixels, with an inner radius set 2 pixels beyond the circular aperture. The background level is highly uncertain due to the uneven distribution of the light from the underlying unresolved stars and adjacent objects, and we estimate uncertainties in resulting colors ~±0.2 mag. Whether these objects qualify as SSCs under the criteria of having $-14.1 < M_V < -11.9$ (O'Connell et al. 1994) is difficult to determine because of the effect the aforementioned uncertainties, along with the contribution of [O III] λλ 4959, 5007 emission to $V$ band (as to opposed to the narrow-$V$ band represented by the F550M filter, which avoids these lines) magnitudes, the effect of nebular continuum emission, and distance uncertainties. The candidates we identify with $M_{F550M} < -10$ mag may however qualify as SSCs under this definition.

In Figure 6 we compare the colors of the YMCs given in Table 3 with the predicted colors of the Bruzual & Charlot (2003) $Z = 0.0004$ spectra used to fit the object spectral energy distributions. A reddening vector based on the mean $A_V$ value for the objects (Table 4) and assuming a Galactic extinction curve is also shown. The colors of the objects roughly match the predictions of the Bruzual & Charlot (2003) models after accounting for the reddening, and indicate typical cluster ages ~10 – 40 Myr. The Bruzual & Charlot models do not include nebular continuum emission, which also likely contributes to the scatter in these values.

## 1. DISCUSSION

The clear evidence of stars ~10 Gyr old in all of the sample objects establishes that they are not protogalaxies, forming their first generation of stars. UCBDs are thus similar to the majority of local BCDs in this respect. BCDs out to $z \cong 0.25$ also appear to be composites of stellar populations of different ages (Barazza et al. 2005). In view of their low stellar masses, the very low metallicities of UCBDs are likely to be the result of the escape of supernova ejecta (see Mac Low & Ferrara 1999; Silich & Tenorio-Tagle 2001; Martin, Kobulnicky, & Heckman 2002), as opposed to youth. The sample objects lie on a extrapolation of the mass / metallicity correlation found for larger SDSS galaxies by Tremonti et al. (2004), which has also been attributed to the supernova-driven escape of enriched gas from low-mass galaxies (see also Erb et al. 2006). Some support for this interpretation is found from the Hα images of BCDs obtained by Gil de Paz et al. (2003), which includes the UCBDs POX 186 and HS 0822+3542.  These images show that in several objects, including these two UCBDs, the Hα emission extends beyond the *R*-band continuum emission of the galaxies, suggesting that "superwinds" similar to those seen in larger disk galaxies (e.g. Lehnert & Heckman 1996) are expelling metal-enriched gas from the objects.  Echelle spectroscopy of the Hα emission of these objects is required to confirm this, as well as high-resolution emission-line imaging with *HST*.  If other extremely metal-poor galaxies such as I Zw 18 and SBS 0335-052W truly lack evolved stars, as reported by Izotov & Thuan (2004) and Izotov, Thuan, & Guseva (2005), respectively, then they would be exceptions to the apparent rule that evolved stars are present in local galaxies of all metallicities.

A high fraction of mass contained in neutral gas is another possible mechanism for the low metal content of these galaxies.  The H I content of UCBDs however shows no clear pattern, with some objects, e.g. SDSS J1152-0040 (Salzer et al. 2002), lacking detectable 21-cm emission, and others having a clear or marginal detection (e.g. SDSS J1119+5130: Huchtmeier, Krishna, & Petrosian 2005, and POX 186: Begum & Chengalur 2005).   A systematic study of the H I properties of the objects in the present sample using observations from the NRAO Greenbank Telescope and Very Large Array will be the subject of a future paper (Hibbard et al., in preparation).

The evidence that the stellar mass in these objects is comprised mainly of stars ~10 Gyr old is consistent with the recent high-resolution hydrodynamical simulations of dwarf galaxy formation in voids (Hoeft et al. 2005; see also Wyithe & Loeb 2006), which include the effect of cosmic reionization at $z \cong 6$ on their baryon content and resulting star formation rates.  These simulations show a significant decline in star formation rate after re-ionization due to photo-evaporation of the baryons in the galaxy cold dark matter halos, which results in low-metallicity dwarf galaxies with predominantly old populations at the present epoch.  Some residual star formation is allowed, which could explain the intermediate-age stars indicated in the objects by our SSP fitting.  The detection of ~10 Gyr old stars in such small galaxies also provides support for the argument of Yan & Windhorst (2004a; see also Yan & Windhorst 2004b) that sub-$L^*$ galaxies dominated the completion of the reionization process at $z \cong 6$.

The origin of the present starbursts in the objects is less clear, but the asymmetric distributions of some

of them,  e.g.  SDSS J1119+5130, suggest that they have been induced by the recent collisions/mergers of

smaller clumps of stars.  The strongest cases for this can be made for POX 186 (Corbin & Vacca 2002) and

HS 0822+3542 (C05 and this study) because of their distorted morphologies and the two-component nature of the latter.  Other objects in the present sample such as SDSS J1127+6410 are however not as dynamically disturbed, while SDSS J1119+5130 and SDSS J1133+6349 have morphologies and burst strengths (as indicated by their colors; Table 1) intermediate between these objects.  This suggests different

stages in a coupled process of merging and star formation, but ACS/HRC observations of more UCBDs are

needed to address this conclusively.  In particular, it is of strong interest to see if ACS/HRC images of

UCBDs similar to HS 0822+3542 resolve into separate clumps of stars as small as components A and B of

this object (Fig. 1).  Corbin & Vacca (2002) and C05 have suggested that such clumps represent galaxy

"building blocks," which have already assembled into larger galaxies in the vast majority of local (and

intermediate-redshift) galaxies.  In the void regions where UCBDs and BCDs reside, consideration of the

timescales associated with the large initial separations of the building blocks (~1 – 10 Mpc; Table 1)

indicates that this assembly process could be delayed by a significant fraction of the Hubble time, allowing

us to witness it only at the current epoch.  Other evidence that the formation of void galaxies is delayed

compared to galaxies in denser environments is found from the high incidence of disturbed galaxies in the

Böotes void (Cruzen, Weistrop, & Hoopes 1997), and statistical evidence from the SDSS that void galaxies have higher specific (mass-normalized) star formation rates than galaxies in richer environments (Rojas et al. 2005; Kauffmann et al. 2004).  As noted by Kauffmann et al. (2004), the dependence of star formation rates on galaxy environment  is similar to their evolution with redshift. UCBDs may represent one extreme of this relationship where we are actually seeing dwarf galaxies in the process of assembling from galaxy building blocks, which at the present epoch contain stars up to ~10 Gyr old.  Support for this interpretation is found from early estimates of the minimum masses for galaxies forming via atomic cooling within cold dark matter halos (White & Rees 1978), which match the stellar masses we estimate for our objects (Table 3).

If UCBDs are the result of collisions between subgalactic clumps containing mostly ~10 Gyr old stars, then the question arises of whether isolated (non-bursting) clumps have been detected in the SDSS and other surveys, and whether they represent a significant population in galaxy voids.  We can partially answer this question by looking at the flux levels of the oldest population components in the SSP model fits (Fig. 2).  The flux levels of these components in the $r$ band correspond to AB magnitudes of approximately 18 in most of the objects.  This is below the $r$-band cut-off magnitude of 17.77 for objects selected for spectroscopy in the SDSS (Strauss et al. 2002), but is above the photometric limit of $r = 22.2$ mag used in the SDSS Second Data Release (Abazajian et al. 2004).  Such a population would thus be detected photometrically but not spectroscopically in the SDSS.  How easily such objects could be distinguished from higher-redshift objects with similar colors remains to be determined, and is beyond the scope of the present study.  The confirmation of such a previously unidentified population would however be very important for understanding galaxy formation, and should be pursued at high priority.

We thank an anonymous referee for constructive suggestions that improved the paper. We also thank Ken Nagamine, Xu Kong, Joe Silk, and Hy Spinrad for helpful discussions, and Abilio Mateus for his help with the SDSS spectra. This research was supported by NASA through grant G0-10180.06-A to at Arizona State University from the Space Telescope Science Institute. The Space Telescope Science Institute is operated by the Association of Universities for Research in Astronomy, Inc., under NASA contract NAS 5-26555.

## APPENDIX A

## CORRECTION FOR RADIAL VARIANCE IN *M/L*

To account for the variance in the *M/L* ratio of the objects outside of the $3''$ diameter region represented by the SDSS spectra, we first randomly selected a sample of 656 low-mass ($< 10^{9.5} \, \mathrm{M}_\odot$) star-forming galaxies from the SDSS, and derived a calibration between their $r - z$ colors within the SDSS fibers and their $M/L_z$ ratios measured from the spectral fitting results with the STARLIGHT code, obtaining

$$\log{(M/L_z)} \cong 1.91 \, (\, r - z \,) \, \text{-} \, 0.80$$

We then converted the F550M – F814W colors of the objects measured outside of the $3''$ SDSS spectroscopic fiber to $r - z$, assuming a simple power-law spectrum, obtaining

$$r - z \, \cong 0.846(\text{F550M} - \text{F814W})$$

The F550M – F814W colors measured outside the apertures were approximately $0.3 - 0.4$ mag redder than those within. This procedure assumes that the aperture centers were coincident with the brightest points of the galaxies as seen in the F550M images.

REFERENCES

Abazajian, K. et al. 2004, AJ, 128, 502

Aller, L.H. 1984, Physics of Thermal Gaseous Nebulae (Dordrecht: Reidel), 102

Aloisi, A., van der Marel, R.P., Mack, J., Leitherer, C., Sirianni, M., Tosi, M. 2005, ApJ, 631, L45

Arp, H. 1965, ApJ, 142, 402

Barazza, F.D. et al. 2005, astro-ph/0512307

Begum, A., & Chengalur, J. 2005, MNRAS, 362, 609

Bingelli, B., Popescu, C.C., & Tammann, G.A. 1993, A&AS, 98, 275

Bruzual, G. & Charlot, S. 2003, MNRAS, 344, 1000

Cairós, L.M., Vilchez, J.M., Gonzalez-Perez, J.N., Iglesias-Paremo, J., & Caon, N. 2001, ApJS, 133, 321

Campos-Aguilar, A., Moles, M. & Masegosa, J. 1993, AJ, 106, 1784

Chabrier, G. 2003, PASP, 115, 763

Cid Fernades, R., Gu, Q., Melnick, J., Terlevich, E., Terlevich, R., Kunth, D., Rodrigues Lacerda, R.,

   Joguet, B. 2004, MNRAS, 355, 273

Cid Fernandes, R., Mateus, A., Sodre Jr., L., Stasinska, G., & Gomes, J.M. 2005, MNRAS, 358, 363

Corbin, M.R. & Vacca, W.D. 2002, ApJ, 581, 1039

Corbin, M.R., Vacca, W.D., Hibbard, J.E., Somerville, R.S., & Windhorst, R.A. 2005, ApJ, 629, L89

   (C05)

Cruzen, S.T., Weistrop, D., & Hoopes, C.G. 1997, AJ, 113, 1983

Doublier, V., Kunth, D., Courbin, F. & Magain, P. 2000, A&A, 353, 887

Erb, D.K., Shapley, A.E., Pettini, M., Steidel, C.C., Reddy, N.A., & Adelberger, K.L. 2006, ApJ, 644, 813

Gil de Paz, A., Madore, B.F., & Pevunova, O. 2003, ApJS, 147, 29

Hoeft, M., Yepes, G., Gottlöber, & Springel, V. 2005, astro-ph/0501304

Huchra, J.P. 1988, The Extragalactic Distance Scale, ASP Conference Series Vol. 4

    (San Francisco: Astronomical Society of the Pacific), 257

Huchtmeier, W.K., Krishna, G., & Petrosian, A. 2005, A&A, 434, 887

Izotov, Y.I. & Thuan, T.X. 2002, ApJ, 567, 875

-------------------------------- 2004, ApJ, 616, 768

Izotov, Y.I., Thuan, T.X., & Guseva, N.G. 2005, ApJ, 632, 210

Johnson, K., Leitherer, C., Vacca, W.D., & Conti, P.S. 2000, AJ, 120, 1273

Kauffmann, G., White, S.D., Heckman, T.M., Ménard, B., Brinchmann, J. , Charlot, S., Tremonti,

    C., & Brinkmann, J. 2004, MNRAS, 353, 713

Kniazev, A.Y. et al. 2000, A&A, 357, 101

Kniazev, A.Y., Grebel, E.K., Hao, L.H., Strauss, M.A., Brinkmann, J., & Fukugita, M. 2003, ApJ,

    593, L73

Kong, X., Charlot, S., Weiss, A. & Cheng, F.Z. 2003, A&A, 403, 877

Kruger, H., Fritze-v. Alvensleben, U., & Loose, H.-H. 1995, A&A, 303, 41

Kunth, D., Maurogordato, S. & Vigroux, L. 1988, A&A, 204, 10

Kunth, D., & Östlin, G. 2000, A&AR, 10, 1

Larsen, S. S. 2004, in The Formation and Evolution of Massive Young Star Clusters, ASP Conference

    Series, Vol. 322, eds. H.J.G.L.M. Lamers, L.J. Smith, & A. Nota. (San Francisco: Astronomical

    Society of the Pacific), 19

Le Borgne, J.-F. et al. 2003, A&A, 402, 433

Lehnert, M.D., & Heckman, T.M. 1996, ApJ, 462, 651

Leitherer, C., & Heckman, T.M. 1995, ApJS, 96, 9

Loose, H.H. & Thuan, T.X. 1985, Star Forming Galaxies and Related Objects, ed. D. Kunth,

T.X. Thuan, & J. Tran Thanh Van (Paris: IAP), 73

Mac Low, M.-M., & Ferrara, A. 1999, ApJ, 513, 142

Martin, C.L., Kobulnicky, H.A., & Heckman, T.M. 2002, ApJ, 574, 663

Mateus, A., Sodre Jr., L., Cid Fernandes, R., Stasinska, G., Schoenell, W., & Gomes, J.M. 2005,

astro-ph/0511578

Mathis, H., Charlot, S., & Brinchmann, J. 2006, MNRAS, 365, 385

Momany, Y. et al. 2005, A&A, 439, 111

Mould, J. 1995, Aust. J. Phys., 48, 1093

Noeske, K., Guseva, N.G., Fricke, K.J., Izotov, Y.I., Papaderos, P. & Thuan, T.X. 2000, A&A, 361, 33

O'Connell, R.W., Gallagher, J.S., & Hunter, D.A. 1994, ApJ, 433, 65

Oey, M.S. 2004, A&SS, 289, 269

Oey, M.S., Watson, A.M., Kern, K., & Walth, G.L. 2005, AJ, 129, 393

Östlin, G. & Mouhcine, M. 2005, A&A, 433, 797

Papaderos, P., Loose, H.-H., Thuan, T.X. & Fricke, K.J. 1996, A&AS, 120, 207

Popescu, C.C., Hopp, U., & Rosa, M.R. 1999, A&A, 350, 414

Pustilnik, S.A., Kniazev, A.Y., Pramskij, A.G., Ugryumov, A.V., & Masegosa, J. 2003,

A&A, 409, 917

Pustilnik, S.A., Kniazev, A.Y. & Pramskij, A.G. 2005, A&A, 443, 91

Raimann, D., Bica, E., Storchi-Bergmann, T., Melnick, J., & Schmitt, H. 2000, MNRAS, 314, 295

Rojas, R.R., Vogeley, M.S., Hoyle, F., & Brinkmann, J. 2005, ApJ, 624, 571

Salzer, J., Rosenberg, J.L., Weisstein, E.W., Mazzarella, J.M., & Bothun, G.D. 2002, AJ, 124, 191

Schechter, P.L. 1980, AJ, 85, 801

Schlegel, D.J., Finkbeiner, D.P., & Davis, M. 1998, ApJ, 500, 525

Searle, L., & Sargent, W.L.W. 1972, ApJ, 173, 25

Silich, S.A. & Tenorio-Tagle, G. 2001, ApJ, 552, 91

Sirianni, M. et al. 2005, PASP, 117, 1049

Spergel, D. et al. 2003, ApJS, 148, 175

Strauss, M. A. et al. 2002, AJ, 124, 1810

Telles, E. & Terlevich, R. 1995, MNRAS, 275, 1

Telles, E., Melnick, J., & Terlevich, R. 1997 MNRAS, 288, 79

Telles, E., & Maddox, S. 2000, MNRAS, 311, 307

Thuan, T.X. 1987, in Nearly Normal Galaxies, ed. S.M. Faber (New York: Springer), 67

Thuan, T.X., Izotov, Y.I., & Lipovetsky, V.A. 1997, ApJ, 477, 661

Tremonti, C.A. et al. 2004, ApJ, 613, 898

van den Bergh, S. 2006, AJ, 131, 304

van Zee, L., Salzer, J.J. & Skillman, E.D. 2001, AJ, 122, 121

Westera, P., Cuisinier, F., Telles, E. & Kehrig, C. 2004, A&A, 423, 133

White, S.M. & Rees, M.J. 1978, MNRAS, 183, 341

Whitmore, B. 2000, Dynamics of Galaxies: From the Early Universe to the Present, ASP Conference

    Series Vol. 197, ed. F. Combes, G.A. Mamon, and V.G. Charmandaris (San Francisco:

    Astronomical Society of the Pacific), 197

Wyithe, J.S. & Loeb, A. 2006 astro-ph/063550

Yan, H. & Windhorst, R.A. 2004a, ApJ, 600, L1

-------------------------------- 2004b, ApJ, 612, L93

TABLE 1

SAMPLE OBJECTS

| SDSS Name | Other Name | $g$ mag. | $z$ | Distance from Milky Way (Mpc) | Distance to Nearest Neighbor (Mpc)[1] |
|---|---|---|---|---|---|
| J013352.56+134209.4 | | 18.1 | 0.0087 | 38.6 | 3.5 |
| J023628.77-005829.7 | SHOC 129 | 18.0 | 0.0083 | 33.0 | 4.0 |
| J024815.93-081716.5 | SHOC 137 | 16.4 | 0.0046 | 20.2 | 0.10 (dwarf) |
| J082555.44+353231.9 | HS 0822+3542 | 17.8 | 0.0024 | 15.4 | 0.02 (dwarf) |
| J111934.35+513012.1 | Arp's Galaxy | 16.9 | 0.0044 | 17.6 | 8.8 |
| J112742.96+641001.5 | SHOC 325 | 17.2 | 0.0078 | 39.4 | 10.4 |
| J113341.19+634925.8 | SHOC 334 | 17.4 | 0.0070 | 35.4 | 10.4 |
| J113543.03+015325.0 | | 17.9 | 0.0053 | 26.9 | 0.47 (dwarf) |
| J115247.51-004007.6 | UM 463 | 17.4 | 0.0046 | 26.9 | 2.4 |

[1]Cases where the nearest neighbor is a dwarf galaxy are noted in parentheses.

TABLE 2

*HUBBLE SPACE TELESCOPE* OBSERVATIONS[1]

| Object | UT Date of Observation | Integration time, s | | |
|---|---|---|---|---|
| | | F330W | F550M | F814W |
| J013352.56+134209.4 | 9/9/04 | 1330 | 2530 | 1030 |
| J023628.77-005829.7 | 9/25/04 | 1638 | 1896 | 1338 |
| J024815.93-081716.5 | 3/11/05 | 1624 | 1688 | 1556 |
| J082555.44+353231.9 | 10/4/04 | 1666 | 1906 | 1366 |
| J111934.35+513012.1 | 11/12/04 | 1702 | 2064 | 1402 |
| J112742.96+641001.5 | 9/24/04 | 1774 | 2044 | 1474 |
| J113341.19+634925.8 | 9/21/04 | 1774 | 2044 | 1474 |
| J113543.03+015325.0 | 9/21/04 | 1474 | 1888 | 1340 |
| J115247.51-004007.6 | 7/26/05 | 1452 | 1964 | 1452 |

[1]Magnitude uncertainties are approximately 0.08 for F550M and 0.03 for F330W and F814W. Details of the measurement of the magnitudes for SDSS J0825+3532 can be found in Corbin et al. (2005).

TABLE 3

PHOTOMETRY FROM *HST* IMAGES[1]

--------------------------------------------------------------------------------------------------------------

| Object | Aperture Diameter, ″ | $\mu_{lim}$(814W) | AB(F550M) | AB(F330M) -AB(F550M) | AB(F550M) -AB(F814W) | $M_{F550M}$ |
|---|---|---|---|---|---|---|
| J013352.56+134209.4 | 5.8 | 21.5 | 19.20 | -0.11 | 0.22 | -13.7 |
| J023628.77-005829.7 | 7.2 | 21.7 | 18.16 | 1.08 | 0.35 | -14.4 |
| J024815.93-081716.5 | 24.7 | 21.9 | 16.67 | -0.58 | 0.31 | -14.9 |
| J082555.44+353231.9 | | 21.7 | | | | |
|    Component A | 1.8 | | 19.55 | -0.46 | -0.07 | -11.4 |
|    Component B | 0.9 | | 21.37 | -0.47 | -0.07 | -9.6 |
| J111934.35+513012.1 | 8.8 | 21.8 | 17.06 | -0.10 | 0.03 | -14.2 |
| J112742.96+641001.5 | 6.9 | 21.9 | 17.37 | 1.13 | 0.35 | -15.6 |
| J113341.19+634925.8 | 5.0 | 21.9 | 18.02 | -0.06 | 0.18 | -14.7 |
| J113543.03+015325.0 | 6.3 | 21.7 | 17.96 | 0.87 | 0.43 | -14.2 |
| J115247.51-004007.6 | 4.2 | 21.9 | 17.97 | 0.14 | 0.14 | -14.2 |

[1]Aperture Diameter represents the major axis length of the ellipse within which flux was measured.

$\mu_{lim}$(814W) represents the limiting surface brightness of the F814W images, in mag arcsec$^{-2}$, as measured from the sky

level. $M_{F550M}$ is the absolute magnitude measured from the AB(F550M) values and calculated using the distances in

Table 1.

TABLE 4

ESTIMATED INTERNAL EXTINCTIONS AND STELLAR MASSES[1]

-----------------------------------------------------------------------------------------------------------------

| Object | $A_V$ mag | log (M*/M$_\odot$) | | |
| --- | --- | --- | --- | --- |
| | | $f(< 7)$ | $f(7-9)$ | $f(> 9)$ |
| J023628.77-005829.7 | 0.19 | | 7.46 | |
| | | 0.01 | 0.24 | 0.75 |
| J024815.93-081716.5 | 0.38 | | >7.07 | |
| | | 0.06 | 0.07 | 0.87 |
| J111934.35+513012.1 | 0.03 | | 7.13 | |
| | | 0.01 | 0.09 | 0.90 |
| J112742.96+641001.5 | 0.11 | | 8.27 | |
| | | < 0.01 | 0.02 | 0.98 |
| J113341.19+634925.8 | 0.17 | | 7.77 | |
| | | 0.01 | 0.02 | 0.97 |
| J113543.03+015325.0 | 0.03 | | 7.07 | |
| | | <0.01 | 0.10 | 0.90 |
| J115247.51-004007.6 | 0.38 | | 7.26 | |
| | | 0.03 | 0.15 | 0.82 |

_________________________________________________________________   _______________

[1] The quantities $f(< 7)$, $f(7\text{-}9)$, and $f(> 9)$ represent the approximate fraction of the mass contained in stars of ages below $10^7$ yr, between $10^{7\text{-}9}$ yr, and above $10^9$ yr, respectively, in the areas of the galaxies covered by the spectroscopic fibers. The spectrum of SDSS J013352.56+134209.4 is too noisy to reliably estimate these values. For SDSS J082555.44+353231.9 see Corbin et al. (2005). For SDSS J024815.93 -081716.3 the values are for the northern starburst region only, placing only a lower limit on the total galaxy mass. Estimated errors in $A_V$ are $< 0.1$ mag, and approximately 0.1 dex in log $M^*/M_\odot$.

TABLE 5

YOUNG MASSIVE STAR CLUSTER CANDIDATES

| Object | Cluster Coordinates[a] | | AB(F550M)[b] | AB(F330W) | AB(F550M) | $M_{\text{F550M}}$ |
|---|---|---|---|---|---|---|
| | $\alpha$ | $\delta$ | | -AB(F550M) | -AB(F814W) | |
| J0133+1342 | 01h 33m | 13° 42′ | | | | |
| | 52.43s | 10.01″ | 23.4 | -0.3 | 0.0 | -9.5 |
| | 52.49s | 9.56″ | 23.9 | 0.0 | -0.3 | -9.0 |
| | 52.50s | 10.36″ | 23.4 | -0.3 | -0.2 | -9.5 |
| | 52.50s | 9.57″ | 23.9 | 0.00 | -0.4 | -9.1 |
| J0236-0058 | 02h 36m | -00° 58′ | | | | |
| | 28.82s | 29.19″ | 22.2 | -0.2 | -0.4 | -10.4 |
| J1119+5130 | 11h 19m | 51° 58′ | | | | |
| | 34.30s | 12.93″ | 21.6 | 0.3 | 0.1 | -9.7 |
| J1127+6410 | 11h 27m | 64° 10′ | | | | |
| | 42.98s | 00.8″ | 22.3 | -0.4 | -0.8 | -10.7 |
| J1133+6349 | 11h 33m | 63° 49′ | | | | |
| | 41.34s | 25.49″ | 22.1 | -0.5 | -0.9 | -10.6 |
| | 41.29s | 24.81″ | 23.0 | -0.6 | -0.5 | -9.8 |
| J1152-0040 | 11h 52m | -00° 40′ | | | | |
| | 47.58s | 07.99″ | 21.3 | -0.68 | 0.0 | -10.8 |

[a]In J2000. Uncertainty is approximately ± 1″ .

[b]Uncertainties are ± 0.1 – 0.2mag

FIGURE CAPTIONS

FIGURE 1. -  Color images of nine ultracompact blue compact dwarf galaxies created from *Hubble Space Telescope* ACS/HRC images in the F330W, F550M, and F814W filters.  The intensity scale is logarithmic, and the orientation is north at the top, and east to the left. The angular scale is noted for each object, along with the corresponding physical scale at the adopted distance to each galaxy (Table 1).

FIGURE 2. -  The results of the spectral synthesis fitting to the SDSS spectra of seven of the nine sample objects (§ 3.3).  The observed spectrum is normalized by the flux in the 4010 Å – 4060 Å interval and is plotted at the top with a thick black line.  Emission lines and bad pixels were masked in the fits and are drawn in turquoise.  The final model spectrum is drawn in red.  The bottom spectra are sums of the constituent simple stellar population spectra in four age bins: $t \leq 10^7$ yr (magenta), $10^7 < t \leq 10^8$ yr (blue), $10^8 < t \leq 10^9$ yr (green) and $t \geq 5 \times 10^9$ yr (black).  The nebular continuum emission spectrum is shown with a dotted turquoise line.  The percentage contributions of these components to the observed flux at 4020 Å are shown in parentheses.  The spectrum of SDSS J0248-0817 is that of the starburst region at the galaxy's northern tip (Fig. 1).

FIGURE 3. - Comparison of the F330W, F550M, and F814W fluxes of SDSS J0133+1342 with Bruzual & Charlot (2003) simple stellar population synthesis models, assuming a Chabrier (2003) initial mass function and a metallicity of $Z = 0.0004$.  The dashed line represents the addition of the 1 Myr and 10 Gyr spectra.

FIGURE 4. - Comparison of the F330W, F550M, and F814W fluxes of the central region of SDSS

J0248+0871 with Bruzual & Charlot (2003) simple stellar population synthesis models (§ 3.3). The dashed line represents the addition of the 1 Gyr and 10 Gyr spectra.

FIGURE 5. - *Left panel*: Fraction of light at 4020 Å from the stellar population components from the spectral synthesis fitting (Fig. 2). *Right panel*: Fraction of mass contained in the different stellar population components (§ 3.4). These values are for the areas covered by the SDSS spectroscopic fibers.

FIGURE 6. - Color-color plot for the Young Massive Star Clusters identified in the sample objects (Table

4). The symbols represent each galaxy containing a cluster and are: SDSS J0133+1342 (filled squares),

SDSS J1133+6349 (mult crosses), SDSS J1119+5130 (open circle), SDSS J1152-0040 (open triangle), and

SDSS J0236-0058 (open diamond). The filled circles and connecting line represent the colors of Bruzual &

Charlot (2003) simple stellar population spectra with a metallicity of $Z = 0.0004$ for the ages shown. A reddening vector estimated from the objects' mean $A_V$ values and assuming a Galactic extinction curve is shown at the lower right.